\documentclass[preprintnumbers,two column, showpacs,amsmath]{revtex4}
\usepackage{graphicx}
\usepackage{dcolumn}
\usepackage{bm}

\begin{document}

\title{Quantum discord dynamics of two qubits in the single-mode cavities}

\author{Chen Wang$^{1}$}
\author{Qing-Hu Chen$^{1,2,}$}\email{qhchen@zju.edu.cn}

\address{
$^{1}$Department of Physics, Zhejiang University, Hangzhou 310027,
P. R. China \\
$^{2}$Center for Statistical and Theoretical Condensed Matter
Physics, Zhejiang Normal University, Jinhua 321004, P. R. China
 }
\date{\today}

\begin{abstract}
The dynamics of the quantum discord  for two identical qubits in
both two independent single-mode cavities and a common single-mode
cavity are discussed. For the initial Bell state with correlated
spins, while the entanglement sudden death can occur, the quantum
discord vanishes only at discrete moments in the independent
cavities and never vanishes in the common cavity. Interestingly,
quantum discord and entanglement show  opposite behaviors in the
common cavity, unlike in the independent cavities. For the initial
Bell state with anti-correlated spins,  quantum discord and
entanglement behave in the same way for both independent cavities
and a common cavity.  It is found that the detunnings always
stabilize the quantum discord.
\end{abstract}


\maketitle

\section{Introduction}

Quantum entanglement, originated from nonlocal quantum correlation,
is fundamental in quantum physics both for understanding the
nonlocality of quantum mechanics~\cite{Einstein} and plays an
important role in almost all efficient protocols for quantum
computations and communications~\cite{Nielsen}. Due to the
unavoidable interaction with the environment, an initially entangled
two-qubit system becomes totally disentangled after evolving for a
finite time. This phenomena is called entanglement sudden death
(ESD)~\cite{yu} and has been recently demonstrated
experimentally~\cite{Almeida}. However, the entanglement may fail to
capture the existence of the quantum correlation in some mixed
separate states, in which the entanglement is considered not a good
measure~\cite{ollivier1,oppenheim1}.

Recently, a new kind of the quantum correlation, quantum discord
(QD) has attracted a lot of attentions~\cite{modi1}. It provides the
alternative route for measurement, which is present even under
separable states~\cite {ollivier1}. The definition of the QD can be
interpreted as the difference of the total quantum information of
the two sub-systems $A$ and $B$ before and after the local operation
on the one of them. The QD has been proved as a good measure of the
non-classical correlations beyond entanglement. Furthermore, the QD
has been indicated as the source to speed up the quantum
computations~\cite{datta1,lanyon1}.

Some works have been devoted to the QD dynamics of two qubits
coupled to  Markovian~\cite{Werlang,Altintas} and
non-Markovian~\cite{Fanchini} environments. The comparisons with
entanglement dynamics have been also performed. However, the
relevant study on two two-level atoms (qubits) coupled to
independent or common single-mode cavities without dissipations has
not been found in the literature, to the best of our knowledge. We
believe that the QD dynamics in these qubit systems  is also of
fundamental interest. In addition, some essential pictures can be
clearly described and unfolded in the framework of the simple model
where the exact solutions are available. Actually, the entanglement
dynamics for two independent Jaynes-Cummings (JC) atoms has been well
studied previously~\cite{Eberly1,Yonac,Ficek,Sainz,chen,Agarwal}.
The ESD was observed obviously from the initial Bell states with
correlated spins. This feature would prevent the application of the
entanglement as basic resource for quantum information processing.
What is the consequence for the   QD in this kind of the qubit
system?. It is just the main topic of the present study.

In the present paper, we will study the QD dynamics for two
identical qubits in both two independent  identical single-mode
cavities and one common  single-mode cavity.  Comparisons with the
corresponding pairwise entanglement, i.e. concurrence, are  also
given. The paper is organized as follows. In Sec. II and III, we
derive the time dependent QD in these two systems if initiated from
two typical Bell states.   In Sec. IV, the results are given and
discussions are made. The conclusion is presented in the last
section.


\section{QD in two identical Jaynes-Cummings atoms}
The Hamiltonian of two identical Jaynes-Cummings atoms is shown as
\begin{eqnarray}  \label{ham:1}
H_{\text{JC}}&=&\frac{\Delta}{2}({\sigma}^{A}_{z}+{\sigma}^{B}_{z}) +{\omega}%
(a^{\dag}a+b^{\dag}b) \\
&&+g(a^{\dag}{\sigma}^{A}_{-}+{\sigma}^{A}_{+}a)+ g(b^{\dag}{\sigma}^{B}_{-}+%
{\sigma}^{B}_{+}b).  \nonumber
\end{eqnarray}
where $\sigma^{A(B)}_{k} (k=x,y,z)$ is the Pauli operator of the atom A(B),
shown as $\sigma_z=|\uparrow{\rangle}{\langle}\uparrow|-|\downarrow{\rangle}{%
\langle}\downarrow|$ and $\sigma_x=|\uparrow{\rangle}{\langle}%
\downarrow|+|\downarrow{\rangle}{\langle}\uparrow|$, with $|\uparrow{\rangle}
(|\downarrow{\rangle})$ the excited (ground) state of the two-level atom, $%
a^{\dag} (b^{\dag})$ and $a (b)$ are the creator and annihilator of the
cavity A (B), respectively, $\Delta$ and $\omega$ are the frequencies of the
atom and the cavity, $g$ is the atom-cavity coupling strength. Here we set $%
\hbar=1$ and the detunning $\delta=\Delta-\omega$.

We first study the evolution of the QD initiated from the Bell state with
 anti-correlated spins, which has the following form
\begin{eqnarray}
|\Psi^{(1)}_{\text{Bell}}{\rangle}=\sin{\alpha}|{\downarrow}{\uparrow}{%
\rangle} +\cos{\alpha}|{\uparrow}{\downarrow}{\rangle}.
\end{eqnarray}
Initially, the vacuum state of the cavity is considered, so the initial
state of the whole system can be written as
\begin{eqnarray}
~ |\Psi(0){\rangle}&=&(\sin{\alpha}|{\downarrow}{\uparrow}{\rangle} +\cos{%
\alpha}|{\uparrow}{\downarrow}{\rangle}){\otimes}|00{\rangle}  \label{f:1} \\
&=&\sin{\alpha}|{\downarrow}{\uparrow}00{\rangle}+ \cos{\alpha}|{\uparrow}{%
\downarrow}00{\rangle}.  \nonumber
\end{eqnarray}
The time dependent wave function can be generally expressed
as~\cite{Eberly1}
\begin{eqnarray}
~ |\Psi(t){\rangle}&=&x_1|{\uparrow}{\downarrow}00{\rangle} +x_2|{\downarrow}%
{\uparrow}00{\rangle}  \label{f:2} \\
&&+x_3|{\downarrow}{\downarrow}10{\rangle} +x_4|{\downarrow}{\downarrow}01{%
\rangle},  \nonumber
\end{eqnarray}
where the coefficients are
\begin{eqnarray}
x_1&=&(Ae^{-i\lambda_{+}t}+Be^{-i\lambda_{-}t})\cos{\alpha} \\
x_2&=&(Ae^{-i\lambda_{+}t}+Be^{-i\lambda_{-}t})\sin{\alpha}  \nonumber \\
x_3&=&C(e^{-i\lambda_{+}t}-e^{-i\lambda_{-}t})\cos{\alpha}  \nonumber \\
x_4&=&C(e^{-i\lambda_{+}t}-e^{-i\lambda_{-}t})\sin{\alpha}.  \nonumber
\end{eqnarray}
with the eigenfrequencies   as
\begin{eqnarray}
~ \lambda_{\pm}=\omega+\frac{\delta}{2}{\pm}\frac{\sqrt{\delta^2+G^2}}{2},
\label{eigen:1}
\end{eqnarray}
here $G=2g$. The auxiliary parameters are shown as
\begin{eqnarray}
~ A&=&\frac{1}{2}(1+\frac{\delta}{\sqrt{\delta^2+G^2}})  \label{aux:1} \\
B&=&\frac{1}{2}(1-\frac{\delta}{\sqrt{\delta^2+G^2}})  \nonumber \\
C&=&\frac{G}{2\sqrt{\delta^2+G^2}}.  \nonumber
\end{eqnarray}

The pairwise density matrix from Eq.~(\ref{f:2}) under the standard
basis
$\{|{\downarrow}{\downarrow}{\rangle},|{\downarrow}{\uparrow}{\rangle}
, |{\uparrow}{\downarrow}{\rangle},|{\uparrow}{\uparrow}{\rangle}\}$
is thus  expressed by tracing the freedoms of the cavities
$\rho_{AB}(t)=\text{Tr}
_{cav}\{\rho(t)\}=\text{Tr}_{cav}\{|\Psi(t){\rangle}{\langle}\Psi(t)|\}$,
\begin{eqnarray}
~ {\rho}_{AB}(t)&=&\frac{1}{4} \left(
\begin{array}{llll}
|x_3|^2+|x_4|^2 & 0 & 0 & 0 \\
0 & |x_2|^2 & x^{*}_1x_2 & 0 \\
0 & x^{*}_2x_1 & |x_1|^2 & 0 \\
0 & 0 & 0 & 0
\end{array}
\right).  \label{rho:1}
\end{eqnarray}

With this density matrix, the routine to derive the QD is formally
given in the Appendix A. The von Neumann entropy for two atoms in
Eq.~(\ref{appsab}) is given by
\begin{eqnarray}
S(A,B)&=&-(|x_3|^2+|x_4|^2)\log(|x_3|^2+|x_4|^2) \\
&&-(|x_1|^2+|x_2|^2)\log(|x_1|^2+|x_2|^2),  \nonumber
\end{eqnarray}
and the sub-system entropies in Eq.~(\ref{apps:2}) are shown as
\begin{eqnarray}
S(A)&=&-(1-|x_1|^2)\log(1-|x_1|^2) \\
&&-|x_1|^2\log|x_1|^2,  \nonumber \\
S(B)&=&-(1-|x_2|^2)\log(1-|x_2|^2) \\
&&-|x_2|^2\log|x_2|^2.  \nonumber
\end{eqnarray}

From the Appendix A, one can find that the expressions of the elements in
Eq.~(\ref{apprho}) are
\begin{eqnarray}
v_+&=&|x_3|^2+|x_4|^2,~v_-=0,~w=|x_2|^2,  \nonumber \\
x&=&|x_1|^2,~y=x_1x^{*}_2,~u=0.
\end{eqnarray}
Moreover,
\begin{eqnarray}
X_{1,+}&=&(|x_3|^2+|x_4|^2)\cos^2{\theta}+|x_2|^2\sin^2{\theta},  \nonumber
\\
X_{1,-}&=&|x_1|^2\cos^2{\theta},  \nonumber \\
|Y_1|^2&=&\frac{\sin^2{\theta}}{4}|x_1|^2|x_2|^2.  \nonumber
\end{eqnarray}
and
\begin{eqnarray}
X_{2,+}&=&(|x_3|^2+|x_4|^2)\sin^2{\theta}+|x_2|^2\cos^2{\theta},  \nonumber
\\
X_{2,-}&=&|x_1|^2\sin^2{\theta},  \nonumber \\
|Y_2|^2&=&\frac{\sin^2{\theta}}{4}|x_1|^2|x_2|^2.  \nonumber
\end{eqnarray}
Therefore these parameters are independent of $\phi$. It follows
that we can search the minimum of the conditional von Neumann
entropy by only varying $ \theta$ in the regime $[0,\pi/2]$.
Following the procedures outlined in Appendix A, we can finally
derive the quantum discord. Since $\alpha$ is limited to $(0,\pi/2)$
, it can be numerically checked  that $ \theta=\pi/4$ corresponds to
the minimum of the conditional entropy in the following
calculations. The minimum of the conditional von Neumann entropy
reads
\begin{eqnarray}
S(A|\Pi^{B})=-\sum_{\epsilon=\pm}{\eta}_{\epsilon}\log{\eta}_{\epsilon},
\end{eqnarray}
where
\begin{eqnarray}
{\eta}_{\pm}=\{1{\pm}[(1-2|x_1|^2)^2+4|x_1x_2|^2]^{1/2}\}/2.
\end{eqnarray}
As a result, the quantum discord is finally given by
\begin{eqnarray}
~ \mathcal{D}&=&-(1-|x_2|^2)\log(1-|x_2|^2)-|x_2|^2\log|x_2|^2  \nonumber
\label{qd:1} \\
&&+(|x_3|^2+|x_4|^2)\log(|x_3|^2+|x_4|^2)  \nonumber \\
&&+(|x_1|^2+|x_2|^2)\log(|x_1|^2+|x_2|^2)  \nonumber \\
&&-\sum_{\epsilon=\pm}{\eta}_{\epsilon}\log{\eta}_{\epsilon}.
\end{eqnarray}
For later use, we also list the expression for concurrence derived
in Ref.~\cite{Eberly1} as
\begin{eqnarray}
C_{AB}(t)=|\sin2\alpha|[1-4C^2\sin^{2}(\sqrt{\delta^2+G^2}t/2)].
\end{eqnarray}


Next, we consider the Bell state with correlated spin as the initial atomic
state, which is
\begin{eqnarray}
|\Psi^{(2)}_{\text{Bell}}{\rangle}=\sin\alpha|{\downarrow}{\downarrow}{%
\rangle} +\cos\alpha|{\uparrow}{\uparrow}{\rangle}.
\end{eqnarray}
Including the initial vacuum cavities, the wave function of the whole system
can be expressed as
\begin{eqnarray}
|\Psi{\rangle}(t)&=&x_1|{\uparrow}{\uparrow}00{\rangle} +x_2|{\downarrow}{%
\downarrow}11{\rangle}+x_3|{\uparrow}{\downarrow}01{\rangle}  \nonumber \\
&&+x_4|{\downarrow}{\uparrow}10{\rangle}+x_5|{\downarrow}{\downarrow}00{%
\rangle},
\end{eqnarray}
where the coefficients are
\begin{eqnarray}
x_1&=&(Ae^{-i\lambda_+t}+Be^{-i\lambda_-t})^2\cos\alpha, \\
x_2&=&AB(e^{-i\lambda_+t}-e^{-i\lambda_-t})^2\cos\alpha,  \nonumber \\
x_3&=&C(e^{-i\lambda_+t}-e^{-i\lambda_-t})(Ae^{-i\lambda_+t}+Be^{-i%
\lambda_-t})\cos\alpha,  \nonumber \\
x_4&=&C(e^{-i\lambda_+t}-e^{-i\lambda_-t})(Ae^{-i\lambda_+t}+Be^{-i%
\lambda_-t})\cos\alpha,  \nonumber \\
x_5&=&\sin\alpha,  \nonumber
\end{eqnarray}
The eigenfrequencies and the auxiliary parameters are the same as
those in  Eqs.~(\ref{eigen:1}) and (\ref{aux:1}). Then under the
standard
basis $\{|{\downarrow}{\downarrow}{\rangle},|{\downarrow}{\uparrow}{\rangle}%
, |{\uparrow}{\downarrow}{\rangle},|{\uparrow}{\uparrow}{\rangle}\}$, the
pairwise density matrix is shown as
\begin{eqnarray}
~ {\rho}_{AB}(t)&=&\left(
\begin{array}{llll}
|x_2|^2+|x_5|^2 & 0 & 0 & x^{*}_1x_5 \\
0 & |x_4|^2 & 0 & 0 \\
0 & 0 & |x_3|^2 & 0 \\
x_1x^{*}_5 & 0 & 0 & |x_1|^2
\end{array}
\right).  \label{rho:2}
\end{eqnarray}
Hence, the joint von Neumann entropy is derive as
\begin{eqnarray}
S(A,B)&=&-|x_3|^2\log|x_3|^2-|x_4|^2\log|x_4|^2 \\
&&-\sum_{\epsilon=\pm}\Omega_{\epsilon}\log\Omega_{\epsilon},  \nonumber
\end{eqnarray}
where
\begin{eqnarray}
\Omega_{\pm}&=&\{(|x_1|^2+|x_2|^2+|x_5|^2) \\
&&{\pm}\sqrt{(|x_2|^2+|x_5|^2-|x_1|^2)^2+4|x_1|^2|x_5|^2}\}/2.  \nonumber
\end{eqnarray}
And the sub-system entropy can be derive as
\begin{eqnarray}
S(A)&=&-(|x_2|^2+|x_4|^2+|x_5|^2)\log(|x_2|^2+|x_4|^2+|x_5|^2)  \nonumber \\
&&-(|x_1|^2+|x_3|^2)\log(|x_1|^2+|x_3|^2),~  \label{s:1} \\
S(B)&=&-(|x_2|^2+|x_3|^2+|x_5|^2)\log(|x_2|^2+|x_3|^2+|x_5|^2)  \nonumber \\
&&-(|x_1|^2+|x_4|^2)\log(|x_1|^2+|x_4|^2).~  \label{s:2}
\end{eqnarray}
Similar to the above Bell state with anti-correlated spins, if we
focus on $ \alpha{\in}(0,\pi/2)$, $\theta=\pi/4$ also corresponds to
the minimum of the conditional von Neumann entropy. Hence, the
minimum of the conditional entropy is given by
\begin{eqnarray}
~ S(A|\Pi^B)=-\sum_{\epsilon=\pm}\eta_{\epsilon}\log{\eta_{\epsilon}},
\label{s:3}
\end{eqnarray}
where
\begin{eqnarray}
\eta_{\pm}=\{1{\pm}\sqrt{(1-2|x_1|^2-2|x_3|^2)^2+4|x_1|^2|x_5|^2}\}/2.
\end{eqnarray}
As a result, the quantum discord can be derived from Eqs.~(\ref{s:1}), (\ref
{s:2}), and (\ref{s:3}) as
\begin{eqnarray}
\mathcal{D}=S(B)-S(A,B)+S(A|\Pi^B).
\end{eqnarray}

The concurrence in this case has been also derived previously~\cite{Eberly1}, 
and is also collected here
$$C_{AB}(t)=\max\{0,f(t)\}$$
\begin{eqnarray}
f(t)&=&[1-4C^2\sin^{2}(\sqrt{\delta^2+G^2}t/2)]\\
&&[|\sin2\alpha|-8C^2\sin^{2}(\sqrt{\delta^2+G^2}t/2)\cos^{2}\alpha].\nonumber
\end{eqnarray}
Specially at resonance ($\delta=0$), the entanglement sudden transition occurs
only for $\alpha<\alpha_c$, where $\alpha_c=\pi/4$.

\section{ QD in two identical qubits in one common single-mode cavity}

The Hamiltonian of  two identical qubits interacting with one common
single-mode cavity reads
\begin{equation}
H_{\textrm{DN2}}=\frac \Delta 2({\sigma }_z^A+{\sigma }_z^B)+{\omega
}(a^{\dag }a+\frac 12)+g\sum_{k=A,B}(a{\sigma }_k^{+}+{\sigma
}_ka^{\dag }). \nonumber
\end{equation}
where $\ $ $a^{\dag }\ $ and $a\ $are the creator and annihilator of
the common cavity. Actually, it is just the $N=2$ Dicke
model~\cite{dicke}. The detunning is also set as
$\delta=\Delta-\omega$.

If the initial atom state is selected as the  Bell  state with
anti-correlated spins, we can obtain the time dependent wavefunction
as
\begin{eqnarray}~\label{atcof:1}
|\Psi(t){\rangle}=x_1|{\uparrow}{\downarrow}0{\rangle}+x_2|{\downarrow}{\uparrow}0{\rangle}+x_3|{\downarrow}{\downarrow}1{\rangle},
\end{eqnarray}
with
\begin{eqnarray}
x_1&=&{\langle}{\uparrow}{\downarrow}0|e^{-iH_{\textrm{DN2}}t}|\Psi^{(1)}_{\text{Bell}}{\rangle},\nonumber\\
x_2&=&{\langle}{\downarrow}{\uparrow}0|e^{-iH_{\textrm{DN2}}t}|\Psi^{(1)}_{\text{Bell}}{\rangle},\nonumber\\
x_3&=&{\langle}{\downarrow}{\downarrow}1|e^{-iH_{\textrm{DN2}}t}|\Psi^{(1)}_{\text{Bell}}{\rangle}.\nonumber
\end{eqnarray}
Then the pairwise density matrix under standard basis
$\{|{\downarrow}{\downarrow}{\rangle},|{\downarrow}{\uparrow}{\rangle},
|{\uparrow}{\downarrow}{\rangle},|{\uparrow}{\uparrow}{\rangle}\}$ is given by
\begin{eqnarray}
~ {\rho}_{AB}(t)&=&\left(
\begin{array}{llll}
|x_3|^2 & 0 & 0 & 0 \\
0 & |x_2|^2 & x^{*}_1x_2 & 0 \\
0 & x_1x^{*}_2 & |x_1|^2 & 0 \\
0 & 0 & 0 & 0
\end{array}
\right).  \label{rho:5}
\end{eqnarray}
For the  resonant case ($\delta=0$), we specify the coefficients of
the wavefunction in Eq.~(\ref{atcof:1}) as
\begin{eqnarray}
x_1&=&\frac{\cos\alpha}{2}(\cos{\sqrt{2}{\lambda}t}+1)
+\frac{\sin\alpha}{2}(\cos{\sqrt{2}{\lambda}t}-1),\nonumber\\
x_2&=&\frac{\cos\alpha}{2}(\cos{\sqrt{2}{\lambda}t}-1)
+\frac{\sin\alpha}{2}(\cos{\sqrt{2}{\lambda}t}+1),\nonumber\\
x_3&=&\frac{-i}{\sqrt{2}}(\cos\alpha+\sin\alpha)\sin{\sqrt{2}{\lambda}t}.\nonumber
\end{eqnarray}
After the numerical checks, we find that $\theta=\pi/4$ corresponds
to the minimum of the conditional entropy at arbitrary time. Hence
the QD is described as
\begin{eqnarray}
~ \mathcal{D}&=&-(1-|x_2|^2)\log(1-|x_2|^2)-|x_2|^2\log|x_2|^2  \nonumber
\label{qd:1} \\
&&+(|x_3|^2)\log(|x_3|^2)+(1-|x_3|^2)\log(1-|x_3|^2)  \nonumber \\
&&-\sum_{\epsilon=\pm}{\eta}_{\epsilon}\log{\eta}_{\epsilon},
\end{eqnarray}
with
\begin{eqnarray}
{\eta}_{\pm}=\{1{\pm}[(1-2|x_1|^2)^2+4|x_1x_2|^2]^{1/2}\}/2.
\end{eqnarray}
Besides, the concurrence of the two atoms can also be given as
\begin{eqnarray}
C_{AB}(t)=2\max\{0,|x_1x_2|\}.
\end{eqnarray}

\begin{figure}[tbp]
\includegraphics[scale=0.50]{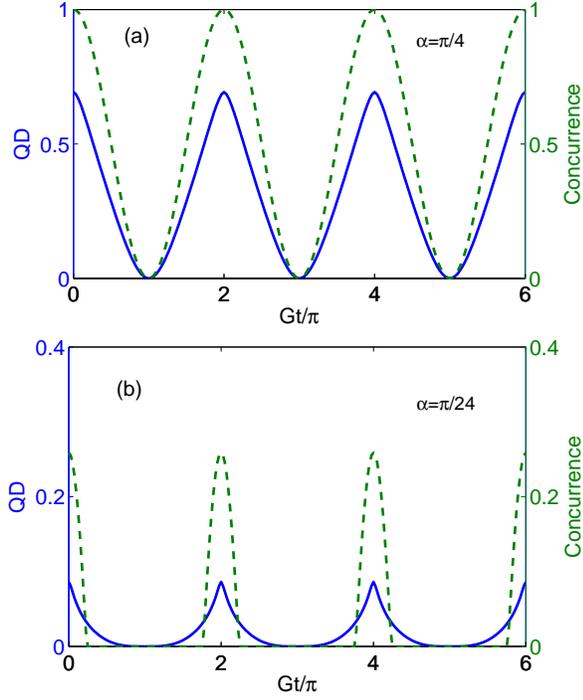}
\caption{(Color online) Resonant dynamics of  QD and
concurrence of  two identical JC atoms with the initial atomic Bell states with anti-correlated
spins   (a) and   correlated
spins   (b). $\omega=1.0$. The blue solid line and the green dashed line are corresponding to
QD and concurrence respectively. } \label{figure1}
\end{figure}

If starting from   the initial  Bell  state with correlated spins,
the wavefunction can be expressed as
\begin{eqnarray}
|\Psi(t){\rangle}&=&x_1|{\uparrow}{\uparrow}0{\rangle}+x_2|{\downarrow}{\downarrow}2{\rangle}
+x_3|{\uparrow}{\downarrow}1{\rangle}+x_4|{\downarrow}{\uparrow}1{\rangle}\nonumber\\
&&+x_5|{\downarrow}{\downarrow}0{\rangle}.
\end{eqnarray}
Hence the pairwise density matrix can be derived as
\begin{eqnarray}
{\rho}_{AB}(t)&=&\left(
\begin{array}{llll}
|x_2|^2+|x_5|^2 & 0 & 0 & x_1^{*}x_5 \\
0 & |x_4|^2 & x^{*}_3x_4 & 0 \\
0 & x_3x^{*}_4 & |x_3|^2 & 0 \\
x_1x_5{*} & 0 & 0 & |x_1|^2
\end{array}
\right).~\label{rho:6}
\end{eqnarray}
The coefficients at resonant condition  are shown as
\begin{eqnarray}
x_1&=&\frac{\cos\alpha}{3}e^{-i{\omega}t}(2+\cos{\sqrt{6}{\lambda}t}),\nonumber\\
x_2&=&\frac{\sqrt{2}\cos\alpha}{3}e^{-i{\omega}t}(\cos{\sqrt{6}{\lambda}t}-1),\nonumber\\
x_3&=&\frac{-i\cos\alpha}{\sqrt{6}}e^{-i{\omega}t}\sin{\sqrt{6}{\lambda}t},\nonumber\\
x_4&=&\frac{-i\cos\alpha}{\sqrt{6}}e^{-i{\omega}t}\sin{\sqrt{6}{\lambda}t},\nonumber\\
x_5&=&\sin\alpha.\nonumber
\end{eqnarray}
Then we can obtain the QD numerically.

While for the concurrence at resonance, we know that the pairwise density
matrix in Eq.~(\ref{rho:6}) has the form as
\begin{eqnarray}
{\rho}_{AB}(t)&=&\left(
\begin{array}{llll}
v_+ & 0 & 0 & u^{*} \\
0 & y & y & 0 \\
0 & y & y & 0 \\
u & 0 & 0 & v_-
\end{array}
\right).~\label{rho:7}
\end{eqnarray}
Then we can derive it as
\begin{eqnarray}
C_{AB}(t)&=&2\max\{0, |x_1x_5|-|x_3|^2,\\
&&|x_3|^2-|x_1|\sqrt{|x_2|^2+|x_5|^2}\}.\nonumber
\end{eqnarray}
From the definition, we find that there exists a critical bound for $\alpha$. The
ESD happens only for $\alpha<\alpha_c$. The
$\alpha_c$ is determined by
\begin{eqnarray}
\tan^{2}\alpha_c-4\tan{\alpha_c}+1=0,
\end{eqnarray}
resulting in $\alpha_c=\arctan(2-\sqrt{3})=\pi/12$.


\begin{figure}[tbp]
\includegraphics[scale=0.50]{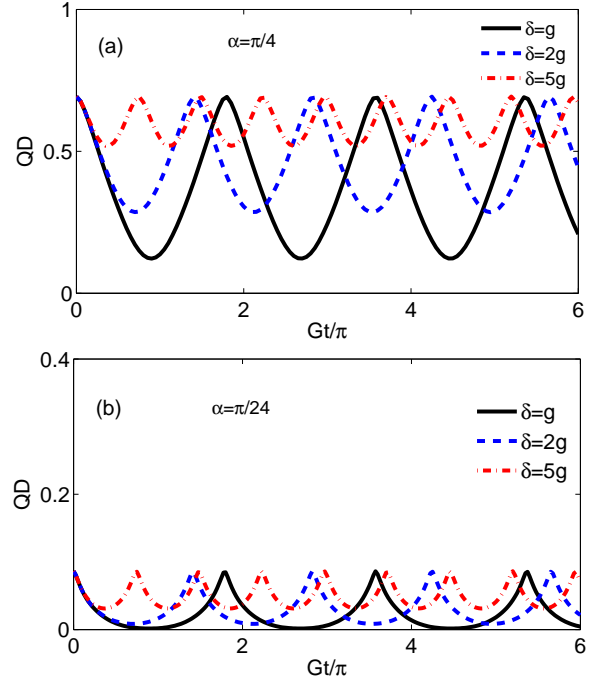}
\caption{(Color online) Off-resonant dynamics of  QD
for two identical JC atoms with the initial atomic Bell states with anti-correlated spins (a)
and correlated spins (b) for different detunnings $\delta=g, 2g$, and
$5g$. $\omega=1.0$.} \label{figure2}
\end{figure}

\section{Results and discussions}

First, we  compare the QD with the concurrence in the two identical
JC atoms with two initial atomic states, i. e. the Bell states with
anti-correlated spins and correlated spins, for zero detunnings. The
results are collected  in Fig. 1. The evolution of both QD and
concurrence for the initial Bell state with anti-correlated spins
display similar behavior, as seen in Fig. 1(a).  Yonac \emph{et al.}~\cite{Eberly1} 
has shown that the ESD only occurs in the initial
atomic Bell states with correlated spin, where the entanglement can
fall abruptly to zero and vanish for a period of time before
revival. It is interesting to note from Fig. 1(b) that during the
period of ESD, QD becomes small but is always finite, except vanish
at discrete moments $t=(2k+1)\pi/G, (k=0,1,2,...$).

Then, we show the effects of the detunnings on the QD in independent
cavities in Fig. 2, starting from these two Bell states.
Interestingly, the amplitude of oscillation of the QD as a function
of time is suppressed monotonically by the detunnings for both
initial Bell stats.  More importantly, the zeros of the QD at
discrete instants shown in Fig. 1(b)  disappear with the finite
detunnings.

\begin{figure}[tbp]
\includegraphics[scale=0.50]{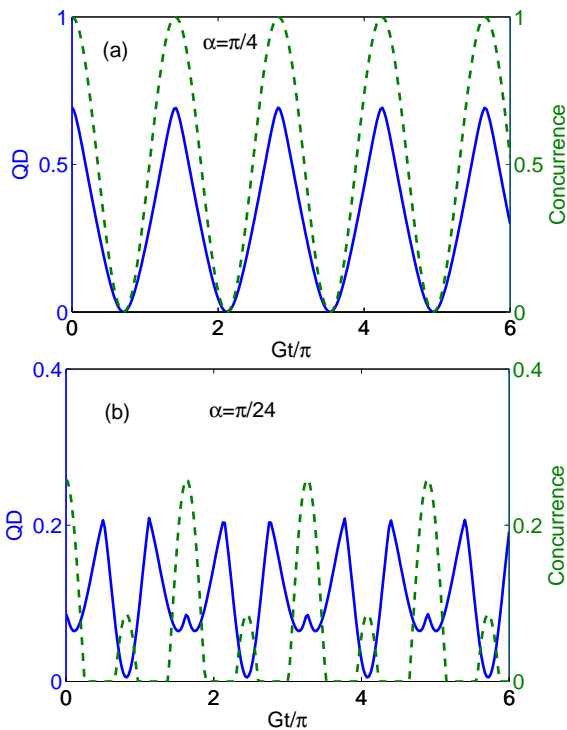}
\caption{(Color online) Resonant dynamics of  QD and
concurrence of  two qubits  coupled to a common cavity with the initial atomic Bell states with anti-correlated
spins   (a) and   correlated
spins   (b). $\omega=1.0$. The blue solid line and the green dashed line are corresponding to
QD and concurrence respectively. }
 \label{figure3}
\end{figure}

Next,  we  compare the QD with the concurrence in the two identical
qubits coupled to the common cavity with two initial atomic Bell
states. The results for zero detunning  are presented in Fig. 3. For
the initial atomic Bell state with anti-correlated spins, similar
behaviors for both QD and concurrence are observed. For the initial
atomic Bell state with correlated spins, one can find from Fig. 3(b)
that  the ESD can occur, but the QD never vanishes. Interestingly,
QD and entanglement show opposite behaviors. Especially, during the
period of ESD, the QD always becomes larger, in sharp contrast with
that observed in the independent cavities (c.f. Fig. 1(b)).

\begin{figure}[tbp]
\includegraphics[scale=0.50]{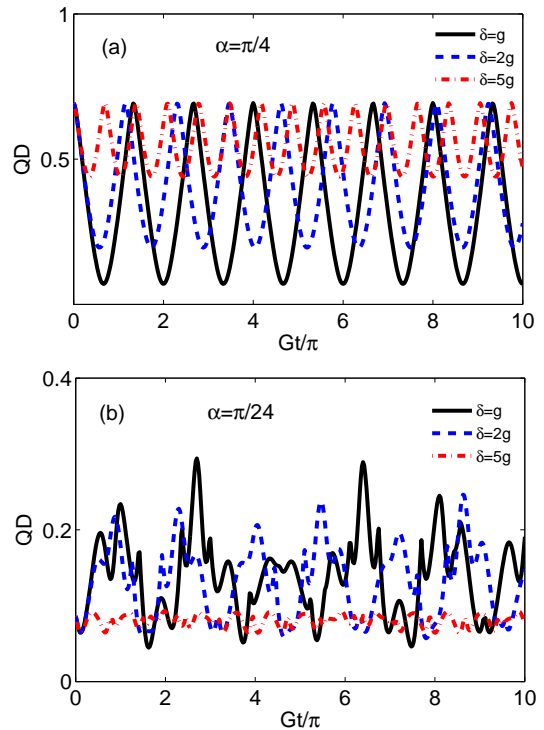}
\caption{(Color online) Off-resonant dynamics of  QD
for two qubits  coupled to a common cavity  with the initial atomic Bell states with anti-correlated spins (a)
and correlated spins (b) for different detunnings $\delta=g, 2g$, and
$5g$. $\omega=1.0$.} \label{figure4}
\end{figure}

Note that the critical parameter $\alpha_c=\pi/12$ below which the
ESD can occurs in the common cavity  is   smaller than
$\alpha_c=\pi/4$ in independent cavities. The instant vanish of QD
is absent in the common cavity, implying that the common cavity
enhance the QD. So, it is suggested  that the quantum correlation in
the common cavity  is stronger than that in independent cavities in
some sense.

The effect of the detunnings on the QD of two qubits coupled to a
common cavity is also studied. As shown in Fig. 4(a),  for the
initial Bell state with anti-correlated spins, the amplitude of
oscillation of the QD as a function of time is also suppressed
monotonically by the detunnings and   little bit larger than that in
two cavities (c.f. Fig. 2(a)). While for the initial atomic  Bell
state with correlated spins with $\alpha=\pi/24$ where ESD can
occur, the oscillation of the QD for two qubits in the common cavity
is suppressed considerably with detunnings, as shown in Fig. 4(b).
Especially, for large detunnings $\delta=5g$, the QD remains almost
unchanged. In this case, we find that the components of the Bell
stats show slightly variation with time for large detunning, due to
the fact that large detunnings prevent the hopping for photons
between different atomic levels to certain degree.

\section{Conclusions}

In this paper, the QD dynamics of two qubits in both independent and
common cavities  are investigated. The comparisons with the
entanglement evolution are also performed.    For the initial atomic
Bell state with anti-correlated spins,  the QD and entanglement show the
similar behaviors for both cavities. But for the initial atomic Bell
state with correlated spins, the QD and entanglement behave in a
remarkably different way. The ESD may occur for both cavities, but
the QD never vanishes suddenly. For the independent cavities, the QD
vanishes only at discrete instants  and can be lifted with finite
detunnings. In   the common cavity  the QD  is always finite. Especially,
the QD and entanglement display an opposite behavior in the common
cavity,  different from those in independent cavities. The detunnings
play important role on the QD dynamics. It always stabilizes the QD,
which could be helpful in the real applications of the QD as the better
resource in quantum information science and quantum computing.

\section{Acknowledgement}

This work was supported by National Natural Science Foundation of China
under Grant Nos. 11174254 and 11104363, National Basic Research Program of
China (Grants No. 2011CBA00103 and No. 2009CB929104).

\appendix

\section{Derivation of the quantum correlation}

In the present paper, the general pairwise density matrix under the standard
basis $\{|{\downarrow }{\uparrow }{\rangle },|{\downarrow }{\uparrow }{%
\rangle },|{\uparrow }{\downarrow }{\rangle },|{\uparrow }{\uparrow }\}$, is
shown as
\begin{eqnarray}~\label{apprho}
~{\rho }_{AB}(t)=\left(
\begin{array}{llll}
v_{+} & 0 & 0 & u^{*} \\
0 & w & y^{*} & 0 \\
0 & y & x & 0 \\
u & 0 & 0 & v_{-}
\end{array}
\right) .
\end{eqnarray}
The von Neumann entropy of the two atoms is
\begin{eqnarray}~\label{appsab}
~S(A,B)=\sum_{\epsilon ={\pm },k=1,2}\Omega _{k,\epsilon }\log \Omega
_{k,\epsilon },
\end{eqnarray}
with
\begin{eqnarray}
\Omega _{1,\pm } &=&\frac{(v_{+}+v_{-}){\pm }\sqrt{(v_{+}-v_{-})^2+4|u|^2}}2,
\nonumber \\
\Omega _{2,\pm } &=&\frac{(w+x){\pm }\sqrt{(w-x)^2+4|y|^2}}2.  \nonumber
\end{eqnarray}
The reduced sub-system density matrices for A and B are obtained as
\begin{eqnarray}
~\rho _A(t) &=&(v_{+}+w)|{\downarrow }{\rangle }_A{\langle }{\downarrow }%
|+(x+v_{-})|{\uparrow }{\rangle }_A{\langle }{\uparrow }|,  \nonumber
 \\
\rho _B(t) &=&(v_{+}+x)|{\downarrow }{\rangle }_B{\langle }{\downarrow }%
|+(w+v_{-})|{\uparrow }{\rangle }_B{\langle }{\uparrow }|.  \nonumber
\end{eqnarray}
Hence, we derive the corresponding von Neumann entropies as
\begin{eqnarray}~\label{apps:2}
S(A) &=&-(v_{+}+w)\log (v_{+}+w)  \nonumber \\
&&-(x+v_{-})\log (x+v_{-}),\\
S(B) &=&-(v_{+}+x)\log (v_{+}+x)  \nonumber \\
&&-(w+v_{-})\log (w+v_{-}),\nonumber
\end{eqnarray}
While for the conditional density matrix $\rho _{A|\Pi ^B}$, projection
basis are considered as
\begin{eqnarray}
~|\Phi _1{\rangle }_B &=&\cos {\theta }|\downarrow {\rangle }_B+e^{i\phi
}\sin {\theta }|\uparrow {\rangle }_B,  \nonumber  \label{appf:1} \\
|\Phi _2{\rangle }_B &=&e^{-i\phi }\sin {\theta }|\downarrow {\rangle }%
_B-\cos {\theta }|\uparrow {\rangle }_B.  \nonumber
\end{eqnarray}
The conditional density operator is expressed as
\[
~\rho _{A|\Pi _k^B}=\Pi _k^B\rho _{AB}\Pi _k^B/p_k,
\]
where $\Pi _k^B=\text{I}_A{\otimes }|\Phi _k{\rangle }_B{\langle }\Phi _k|$
and $p_k=\text{Tr}_{AB}\{\rho _{A|\Pi _k^B}\}$. Specifically under the
projections in Eq.~(\ref{appf:1}),
\begin{eqnarray}
\rho _{A|\Pi _k^B} &=&|\Phi _k{\rangle }_B{\langle }\Phi _k|{\otimes }\{|{%
\downarrow }{\rangle }_A{\langle }\downarrow |X_{k,+}+|{\uparrow }{\rangle }%
_A{\langle }\uparrow |X_{k,-}  \nonumber \\
&&+|{\downarrow }{\rangle }_A{\langle }\uparrow |Y_k+|{\uparrow }{\rangle }_A%
{\langle }\uparrow |Y_k^{*}\}/p_k.  \nonumber
\end{eqnarray}
For $k=1$, we show
\begin{eqnarray}
X_{1,+} &=&v_{+}\cos ^2{\theta }+w\sin ^2{\theta },  \nonumber \\
X_{1,-} &=&x\cos ^\theta +v_{-}\sin ^2{\theta },  \nonumber \\
Y_1 &=&(y^{*}e^{-i\phi }+u^{*}e^{i\phi })\sin {\theta }\cos {\theta }.
\nonumber
\end{eqnarray}
For $k=2$,
\begin{eqnarray}
X_{2,+} &=&v_{+}\sin ^2{\theta }+w\cos ^2{\theta },  \nonumber \\
X_{2,-} &=&x\sin ^2{\theta }+v_{-}\cos ^2{\theta },  \nonumber \\
Y_2 &=&-(y^{*}e^{-i\phi }+u^{*}e^{i\phi })\sin {\theta }\cos {\theta }.
\nonumber
\end{eqnarray}
Then the eigenvalues of the conditional density matrix
reads
\begin{eqnarray}
\eta _{k,\pm } &=&\frac 1{2p_k}\{(X_{k,+}+X_{k,-}){\pm }[(X_{k,+}-X_{k,-})^2
\nonumber \\
&&+4|Y_k|^2]^{1/2}\}.
\end{eqnarray}
The conditional von Neumann entropy is described as
\begin{eqnarray}
S(A|\Pi ^B) &=&\sum_{k=1,2}-p_k\text{Tr}_A\{{\rho _{A|\Pi _k^B}}\log {\rho
_{A|\Pi _k^B}}\} \\
&=&-\sum_{\epsilon =\pm }\sum_{k=1,2}p_k{\eta }_{k,\epsilon }(\theta ,\phi
)\log {{\eta }_{k,\epsilon }(\theta ,\phi )}.  \nonumber
\end{eqnarray}
As a result, the quantum discord can be obtained by
\[
\mathcal{D}=\min \{S(B)-S(A,B)+S(A|\Pi ^B)\},
\]
and the classical correlation is given as
\[
\mathcal{C}=\max \{S(A)-S(A|\Pi ^B)\}.
\]


\begin{references}


\bibitem{Einstein} A. Einstein, B. Podolsky, and N. Rosen, Phys. Rev. \textbf{47}, 777 (1935);
J. S. Bell, Physics \textbf{1}, 195 (1964).


\bibitem{Nielsen} M. A. Nielsen and I. L. Chuan, \emph{Quantum Computational and Quantum information}
(Cambridge University Press, Cambridge, UK 2000).


\bibitem{yu} T. Yu and J. H. Eberly, Phys. Rev. Lett. \textbf{93}, 140404 (2004); T. Yu and J. H. Eberly, 
Phys. Rev. Lett. \textbf{97}, 140403 (2006).

\bibitem{Almeida} M. P. Almeida, F. de Melo, M. Hor-Meyll, A. Salles, S. P. Walborn,
P. H. Souto Ribeiro, and L. Davidovich, Science \textbf{ 316}, 579 (2007).


\bibitem{ollivier1} H. Ollivier and W. H. Zurek, Phys. Rev. Lett. \textbf{88}, 017901 (2001).

\bibitem{oppenheim1} J. Oppenheim, M. Horodecki, P. Horodecki, and R. Horodecki, Phys. Rev. Lett.
\textbf{89}, 180402 (2002).

\bibitem{modi1} K. Modi, A. Brodutch, H. Cable, T. Paterek, and V. Vedral, arXiv:1112.6238 (2011).

\bibitem{lanyon1} B. P. Lanyon, M. Barbieri, M. P. Almeida, and A. G. White, Phys. Rev. Lett.
\textbf{101}, 200501 (2008).

\bibitem{datta1} A. Datta, A. Shaji, and C. M. Caves, Phys. Rev. Lett. \textbf{100}, 050502 (2008).


\bibitem{Werlang} T. Werlang, S. Souza,  F. F. Fanchini,  and C. J. Villas
Boas,  Phys. Rev. A \textbf{80},  024103 (2009).

\bibitem{Altintas} F. Altintas and R.  Eryigit, J. Phys. B: At. Mol. Opt. Phys. \textbf{44},  125501(2011).

\bibitem{Fanchini} F. F. Fanchini,  T. Werlang, C. A. Brasil,  L. G. E. Arruda,
 and A. O. Caldeira,  Phys. Rev. A \textbf{81},  052107 (2010).


\bibitem{Eberly1} M.  Y\"{o}nac, T. Yu and J. H. Eberly, J. Phys. B: At. Mol. Opt. Phys. \textbf{39},
S621 (2006).
\bibitem{Yonac}  M.  Y\"{o}nac, T. Yu and J. H. Eberly,
J. Phys. B: At. Mol. Opt. Phys. \textbf{40}, S45 (2007).

\bibitem{Ficek} S. Chan, M. D. Reid,  and Z. Ficek,  J. Phys. B: At. Mol. Opt. Phys. \textbf{42},
065507 (2009).

\bibitem{Sainz} I. Sainz and G.  Bjork, Phys. Rev. A \textbf{76}, 042313 (2007).


\bibitem{chen}   Q. H. Chen, Y. Yang, T. Liu,  and K. L. Wang,   Phys. Rev. A
\textbf{82}, 052306 (2010).
\bibitem{Agarwal} S. Agarwal, S.M. Hashemi Rafsanjani and J.H. Eberly, arXiv:1201.2928 (2012).

\bibitem{dicke} R. H. Dicke, Phys. Rev. \textbf{ 93}, 99 (1954).

\end{references}
\end{document}